\documentclass[twocolumn,showpacs,preprintnumbers]{revtex4}

\usepackage{amssymb}
\usepackage{amsmath}
\usepackage{graphicx}
\usepackage{dcolumn}
\usepackage{bm}
\usepackage{color}

\begin{document}

\title{Zeno and anti-Zeno effect in an open quantum system in the ultrastrong-coupling regime}
\author{Shu He$^{1}$, Qing-Hu Chen$^{1,2,*}$, and Hang Zheng$^{3}$}
\address{
$^{1}$ Department of Physics, Zhejiang University, Hangzhou 310027,  China \\
$^{2}$  Collaborative Innovation Center of Advanced Microstructures, Nanjing 210093, China \\
$^{3}$ Department of Physics, Shanghai Jiao Tong University, Shanghai 200240, China
 }\date{\today }

\begin{abstract}
We study the quantum Zeno effect (QZE) and quantum anti-Zeno effect (QAZE) of  a two-level system  interacting with an environment of harmonic oscillators, the spin-boson model. By applying a numerically exact method based on matrix product states, the previously obtained picture on the influence of counter-rotating terms has to be modified: For physical bath initial states, the transition from QZE to QAZE with increasing measurement interval is only absent at weak coupling, while present at strong coupling. Furthermore, we find that the transition occurs always for the widely used bare bath initial state. Within a more realistic measurement scheme where only the qubit is projectively measured, the above scenario for the bare bath initial state remains qualitatively unchanged, apart from accelerated decay for intermediate measurement intervals.
\end{abstract}

\maketitle

\section{Introduction}

Frequent measurements of a quantum system may  slow down
its dynamic evolution. This phenomenon is known as the quantum Zeno effect (QZE)~\cite%
{Kofman}. On the other hand,  the decay
rate could also be  accelerated by frequent measurements
under somewhat different conditions. This  opposite effect is generally called the
quantum anti-Zeno effect (QAZE) ~\cite{Facchi}. The change in the evolution
implies a potential strategy to control the quantum dynamics of a target
system. It can be used to protect quantum information ~\cite{Barenco},
suppress decoherence ~\cite{Beige} and even cool down and purify a quantum
system ~\cite{Erez}. Experimentally, both QZE and QAZE were
investigated in various contexts such as trapped ions ~\cite{Itano},
ultracold atoms ~\cite{Streed}, nanomechanical oscillators ~\cite{Chen},
superconducting circuits \cite{Barone1}, and cavity quantum electrodynamics
systems ~\cite{Barone2}. One of the important concepts employed for quantum
control is the QZE-QAZE transition. It occurs in a qubit interacting with
its environment where repetitive measurements are performed which project
the whole system (qubit plus environment) to its initial state. Recently,
the ultra-strong coupling regime between qubit and environment has been
realized within cavity quantum electrodynamics and superconducting circuits ~%
\cite{Bourassa,Abdumalikov,Forn}. In this case, the conventional
rotating-wave approximation (RWA), the standard in quantum optics, is no longer
valid. It is therefore mandatory to go beyond it also in the study of the
Zeno effects.

It is well known that both QZE and QAZE always happen within an RWA-based
approach~\cite{Kofman,RWA}. However, recent
studies going beyond the RWA have shown that the
counter-rotating terms  play a significant role for both QZE and QAZE ~%
\cite{Zheng08,XCao,suncp}, although they were only
partially taken into account. Zheng et al. found that there is no QAZE for hydrogen \cite%
{Zheng08} if the decay from a physical excited initial state is computed
including counter-rotating terms to the second order. Ai et al. studied
systematically the spontaneous decay phenomenon of a two-level system under
the influence of both the environment and repetitive measurements without
the RWA~\cite{suncp}, and found that the QAZE can
still happen in some cases ( even without the RWA
) if the initial state is a bare excited state. For a physically excited
state, they also observed that the QAZE disappears and the QZE is always
robust.

For a reliable study of the Zeno effects in quantum dissipative systems, the
exact dynamics is needed in principle. Actually, the dynamics of quantum
open systems has been studied for a long time using a variety of methods; an
incomplete list is given by the authors of Refs.~\cite%
{Tani,Shao,Yan,wanghb,Prior1,caojs,zhaoy,maj,wangc,Prior2,Chin2,kast,Friend}%
. For a Lorentzian spectral function, it is generally believed that the
numerically exact hierarchical equations-of-motion approach can be
applied. Exact dynamics for more general baths, such as Ohmic and sub-Ohmic
baths, were claimed by several groups ~\cite{wanghb,kast,Friend}. For
the Zeno effects, the other crucial ingredient is the measurement scheme. In
most experimentally realizable processes, measurements are applied on the
system (qubit) only and the state of the environment after each measurement
will deviate from its initial state ~\cite{Gordon}. This applies certainly
in the ultra-strong coupling regime due to the strong correlation between
the system and its environment. The ideal assumption that the measurement
projects the state of the system together with the environment to the total
initial state breaks down. Recently, the effect induced by partial
measurement was considered in the dephasing process based on an exactly
solvable model ~\cite{Chaudhry}. Therefore it is very important to see how
this more realistic measurement scheme affects the QZE-QAZE transition of
the dissipative open quantum system in the ultra-strong coupling regime.

In this paper, we study the QZE and QAZE of an open quantum system where a
two-level system interacts with its environment for both Ohmic and sub-Ohmic
baths, the paradigmatic spin-boson model. We focus mainly on the
ultra-strong coupling regime where both the RWA and the
ideal measurement assumption are not valid. We  employ a highly
efficient and numerically exact method based on a recently proposed
time-dependent variational principle (TDVP) for matrix product states (MPS)~%
\cite{Haegeman}, improved in~\cite{Friend,Zach}, to study the evolution from
several initial states. It was shown that quantum fluctuations of the
environment manifested in a huge number of environmental bosons generated
during time evolution can be precisely described within this approach~\cite%
{Friend}. We  compare our result of the QZE and QAZE transitions with
previous studies based on a conventional unitary transformation (UT)
approach, where the effect of counter-rotating terms is not fully taken into
account ~\cite{Zheng08,suncp}. By TDVP, we  also study the QZE and QAZE
within the more realistic measurement scheme described above, where only the
qubit is projected onto the initial state while the environment continues to
evolve~\cite{he_phd}.

The paper is organized as follows. In Sec. II, we briefly introduce the
TDVP approach as applied to the open quantum system. Then we employ it to
study the spin dynamics, and compare to the results obtained by the UT
approach in Sec. III. In Sec. IV, we use the numerical exact TDVP method to
study the QZE. The results for two initial states are given,
and comparison to previous results by the UT approach is performed.
Furthermore, the TDVP method is used to study the QZE and QAZE within a more
realistic measurement scheme. We close with a short summary in Sec. V.

\section{Time-dependent variational principle}

A qubit coupled to its environment of harmonic oscillators can be described
by the following spin-boson Hamiltonian ($\hbar =1$):
\begin{equation}
H=\frac{1}{2}\Delta \sigma _{z}+\frac{1}{2}\sum_{k}g_{k}(a_{k}^{\dag
}+a_{k})\sigma _{x}+\sum_{k}\omega _{k}a_{k}^{\dag }a_{k},  \label{Hamiorig}
\end{equation}%
where the qubit has an energy splitting of $\Delta $ and coupling strength $%
g_{k}$ to boson modes of frequency $\omega _{k}$. Since MPS-based methods
work particularly well on one-dimensional chain models with short range interaction.
Hamiltonian (\ref{Hamiorig}) is transformed into a representation of a one-dimensional
semi-infinite chain with nearest interaction through an orthogonal
polynomial mapping(see details in Ref.~\cite{Plenio}):
\begin{widetext}
\begin{equation}
H_{\text{chain}}=\frac{1}{2}\Delta \sigma _{z}+\frac{\sigma _{x}}{2}%
c_{0}(b_{0}+b_{0}^{\dag })+\sum_{k=0}^{L}[\epsilon _{k}b_{k}^{\dag
}b_{k}+t_{k}(b_{k}^{\dag }b_{k+1}+b_{k+1}^{\dag }b_{k})],  \label{Hamitrans}
\end{equation}
\end{widetext}
where $b_{k}^{\dag }$($b_{k}$) are creation(annihilation) operators for
transformed new boson modes with $\epsilon _{k}$ describing their frequency
and $t_{k}$ representing the internal nearest coupling strength. $c_{0}$
characterizes the effective coupling between the system and new effective
environment. $t_{k},\epsilon _{k}$ and $c_{0}$ are determined by the
specific form of the spectral function $J(\omega )=\sum_{k}g_{k}^{2}\delta
(\omega -\omega _{k})$. The truncation site number $L$ is set to be large
enough to ensure convergence.

Instead of using standard matrix product representation with fixed local
eigenbasis of boson modes $|n_{k}\rangle $(with truncation number $d_{k}$),
an optimized boson basis  $|\tilde{n}_{k}\rangle $(with truncation
number $d_{O,k}\ll d_{k}$ ) is further employed through an additional
isometric map $V_{\tilde{n}_{k},n_{k}}$, which was introduced in Ref.~\cite%
{guo2012critical} to study the quantum criticality of the spin-boson model.
In the case of a large variance of photon number, this mapping allows
significant compression of local boson basis and dramatically enlarges the
maximal photon number attainable during dynamical evolution.

The TDVP introduced in Ref. ~\cite{Haegeman} is employed to calculate the
evolution here. It is based on the Dirac-Frenkel variational principle by
projecting the Schr\"odinger equation onto the tangent space of the MPS
manifold. The obtained optimal equation shares a similar form with the
original MPS algorithm based on the Suzuki-Trotter decomposition \cite%
{Suzuki}, evolving the system site by site in small time steps $\delta t$. A
further improvement was given in Ref. ~\cite{Friend} to incorporate   the optimized boson basis
in the equation of motion. It was proved that TDVP is
equivalent to a Lie-Trotter splitting approach ~\cite{MSuzuki}, where one
evolves each integrable part of the Schr\"odinger equation in the MPS
framework. Unlike the Suzuki-Trotter decomposition of the time-evolution
operator $U(t)=e^{-iHt}$, errors of TDVP arise only in the integration
scheme.

We outline briefly the approach. The MPS algorithm works well on any one-dimensional
chain model (~\ref{Hamitrans}), the optimized boson basis  is added to the original MPS network to
increase the maximal photon number in the Fock basis. Then TDVP is employed
to calculate the evolution. The error of the whole algorithm comes from the
following procedures. First, MPS-bond dimension $D$ bounds the maximal
entanglement of the state that can be described in current MPS subspace.
Second, the truncation numbers, $d_{O,_{k}}$, of the local basis $d_{k}$ and
the optimized boson basis restrict the maximal attainable photon numbers and therefore the
environmental quantum fluctuations. Finally, time step $\delta t$ introduces
an error of integration in the equation of motion. All of these errors can
be well controlled by setting the parameters adequately large (for $%
D,d_{k},d_{O,k}$), or small (for $\delta t$) to obtain a
numerically reliable result. Unless otherwise specified, we set $D=6$, $%
d_{k} =40$, $d_{O,k}=12\sim20$ and $\delta t = 0.1\sim0.4$ for all of the
calculations in this work. In addition, to avoid the reflection of the
evolution from the end of the chain, we generally choose the size of the
chain in the Hamiltonian $(\ref{Hamitrans})$ $L = \frac{2}{3}\omega_c T$ (T is the simulated time range).
All results in the rest of the paper are carefully checked to reach
convergence.

In this work, we focus on the power-law spectral function which can be
written as:
\begin{equation*}
J(\omega )=\sum_{k}g_{k}^{2}\delta (\omega -\omega _{k})=2\alpha \omega
^{s}\omega _{c}^{1-s}\Theta (\omega _{c}-\omega ),
\end{equation*}%
where $\alpha $ is the dimensionless coupling strength, $\omega _{c}$ is the
maximal frequency of the environment, and $\Theta (x)$ is the step function. The
bath exponent $s$ classifies the reservoir into super-Ohmic $\left(
s>1\right) $, Ohmic $\left( s=1\right) $, and sub-Ohmic $\left( s<1\right) $
types respectively. We set  $\omega_c$ as the energy unit. For the data presented below, we typically
choose $\Delta/\omega_c = 0.1$. All quantities  plotted throughout this paper  are dimensionless.

\section{Spin dynamics}

We first study the spin dynamics and compare the result of the TDVP method with
previous analytical studies based on a conventional unitary transformation
approach proposed in Ref. ~\cite{Zheng}.

The basic idea of the UT approach is to perform a polaronic transformation
with a certain shift $\lambda _{k}\ $for each boson mode to the rotated
spin-boson Hamiltonian,%
\begin{equation}
H^{\prime }=\exp (S)H\exp (-S),  \label{transformed}
\end{equation}%
where%
\begin{equation*}
S=\sum_{k}\lambda _{k}\left( a_{k}^{\dag }-a_{k}\right) \sigma _{x},\
\lambda _{k}=\frac{g_{k}}{2\omega _{k}}\xi _{k},
\end{equation*}%
followed by dropping the counter-rotating terms after the transformation.
The shift $\lambda _{k}\ $are determined variationally by minimizing the
expectation value of the higher order terms on the ground state of the
transformed Hamiltonian. The transformed Hamiltonian is of the RWA type~\cite{Zheng}, which thus can be used as a versatile
platform to study quantum phase transitions, nonequilibrium dynamics and
Zeno effects in the spin-Boson model ~\cite{Zheng08,XCao,Tong, zheng2} as
straightforwardly as in the original RWA framework.

In the previous UT study ~\cite{Zheng}, the ground state of the transformed
Hamiltonian $H^{\prime }$ in the RWA form
is easily obtained as
\mbox{$|GS\rangle ^{\prime }=|\downarrow \rangle \otimes
|\{0\}_{k}\rangle $}. The dynamics can then be straightforwardly studied for
an initial state
\begin{equation}
|\psi (t=0)\rangle ^{\prime }=\frac{1}{\sqrt{2}}\left( 1+\sigma _{x}\right)
|GS\rangle ^{\prime }  \label{initial_old}
\end{equation}%
It takes the following form for the original Hamiltonian (\ref{Hamiorig})
\begin{equation}
|\psi (t=0)\rangle =\frac{1}{\sqrt{2}}\left( 1+\sigma _{x}\right) \exp \left[
-S\right] |\downarrow \rangle |\{0\}_{k}\rangle   \label{initial_3}
\end{equation}%
This initial state is the same as in Eq.(6) of Ref.~\cite{Tong} for the
spin-boson Hamiltonian.

\begin{figure}[htp]
\includegraphics[scale=0.6]{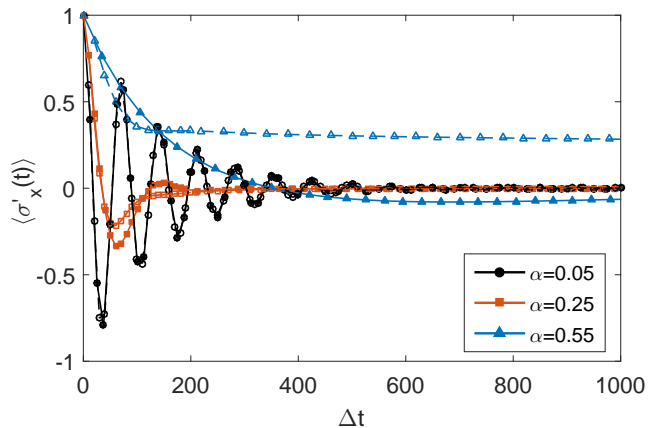}
\caption{(Color online) Evolution of the coherence $\langle \protect\sigma%
_x^\prime(t)\rangle$ for the Hamiltonian (\protect\ref{transformed}) from
the initial state Eq. (\protect\ref{initial_old}). Lines with solid symbols denote the
results from TDVP while those with open symbols are results of UT. Both methods show good
agreement only at weak coupling ($\protect\alpha = 0.05$). }
\label{Fig0}
\end{figure}

The spin dynamics $\langle \sigma _{x}^{\prime }(t)\rangle$ in the framework
of Hamiltonian (\ref{transformed}) can be straightforwardly calculated by UT
approaches. To show the effectiveness of the UT approach, we also use TDVP
to calculate $\langle \sigma _{x}^{\prime }(t)\rangle$ by the evolution of
initial states (\ref{initial_3}) to Hamiltonian (\ref{Hamiorig}). As shown
in Fig. \ref{Fig0}, both methods give an oscillatory decay for the coherence
$\langle \sigma _{x}^{\prime }(t)\rangle$ and agree well for a long period
of time in the weak coupling regime ($\alpha =0.05$) of the Ohmic reservoir.
When the coupling strength increases, the UT results obviously deviate from
the TDVP ones. It follows that the UT approach is only suitable in the weak
coupling regime, where some analytical results are 
available. However, as the coupling strength increases, it becomes obvious
that the UT approach does not describe the correct dynamics. The significant differences
between RWA and non-RWA in the strong coupling regime have been  reported in Ref. (~\cite{Prior1}). The newly
developed TDVP approach would be a better candidate. In the next section, we
 generalize the TDVP approach to study the QZE and QAZE.

\section{QZE and QAZE transition in open quantum system}

With this efficient numerical technique, we turn to the QZE and QAZE in open
quantum system described by Hamiltonian (\ref{Hamiorig}). To study the
QZE-QAZE transition we focus on the survival probability $P_{s}(t=n\tau )$ ~%
\cite{Kofman}, defined as the probability of finding initial state after $N$
successive measurements with equal time interval $\tau \ $(zero \
temperature):
\begin{equation*}
P_{s}(t=n\tau )=|\langle \psi (0)|\exp (-iH\tau )|\psi (0)\rangle |^{2n}
\end{equation*}

If the time interval $\tau $ is short enough, we can further simplify it:
\begin{equation}
P_{s}(t=n\tau )\approx \big[1-(\langle H^{2}\rangle -\langle H\rangle
^{2})\tau ^{2}\big]^{n}\approx \exp (-\gamma (\tau )t).  \label{Psaprox}
\end{equation}%
where $\gamma (\tau )$ is the effective decay rate induced by measurements.
The exponential form of $P_{s}(\tau )$ in Eq. (\ref{Psaprox}) is valid under
certain conditions. First, $\tau $ should be shorter than the typical time
scale of the evolution without measurement. This can be fulfilled by
limiting the time scale in the calculation to be shorter than the time scale
of the bare Rabi oscillation, namely $t\leq \pi /\Delta $. Moreover, the
correlation between qubit and environment should be weak so that the
environment experiences no significant evolution during the interval of
repetitive measurements leading to the assumption that the whole system
(qubit plus its environment) collapses to its initial state after each
measurement. This assumption, although commonly employed in previous
studies, becomes invalid when the system and its environment is strongly
correlated due to the strong coupling between them. This system-environment
correlation effects will affect the properties of QZE-QAZE transition. To  compare with
previous studies, we  present our results also under this assumption
first, then discuss the Zeno effect beyond this assumption, i. e. if only
the qubit is measured.

In the study of Zeno effects, two initial states are widely used in the
literature. One is the product state of the photon vacuum and the atomic
excited state~\cite{Kofman,RWA},
\begin{equation}
|\psi _{1}(t=0)\rangle =|\uparrow \rangle \otimes |\{0\}_{k}\rangle .
\label{initial_1}
\end{equation}%
which we denote as the bare bath initial state. In the UT approach~%
\cite{Zheng08,XCao}, the initial state is usually chosen as
\begin{equation}
|\psi _{2}(t=0)\rangle =\exp [-S]\biggl(|\uparrow \rangle \otimes
|\{0\}_{k}\rangle \biggr).  \label{initial_2}
\end{equation}%
Note that $|\uparrow \rangle \otimes |\{0\}_{k}\rangle $ is obtained by
flipping the spin of the ground state of the transformed Hamiltonian (\ref%
{transformed}) in the RWA form, and thus we call
state (\ref{initial_2}) the physical bath initial state. It is an
entangled state of the system and the environment, and just corresponds to the
state of the measured system prepared in an eigenstate of the measured
observable. On the Bloch sphere form of qubit, one can write it as
\begin{equation}
|\psi _{2}(t=0)\rangle =\cosh (B)|\{0\}_{k}\rangle |\uparrow \rangle -\sinh
(B)|\{0\}_{k}\rangle |\downarrow \rangle ,
\end{equation}%
\ where $B=\sum_{k}\lambda _{k}\left( a_{k}^{\dag }-a_{k}\right) $. Both
initial states (\ref{initial_1}) and (\ref{initial_2}) will be employed to
study the Zeno effects in this paper.

\subsection{Measuring the whole system}

\begin{figure}[htp]
\includegraphics[scale=0.6]{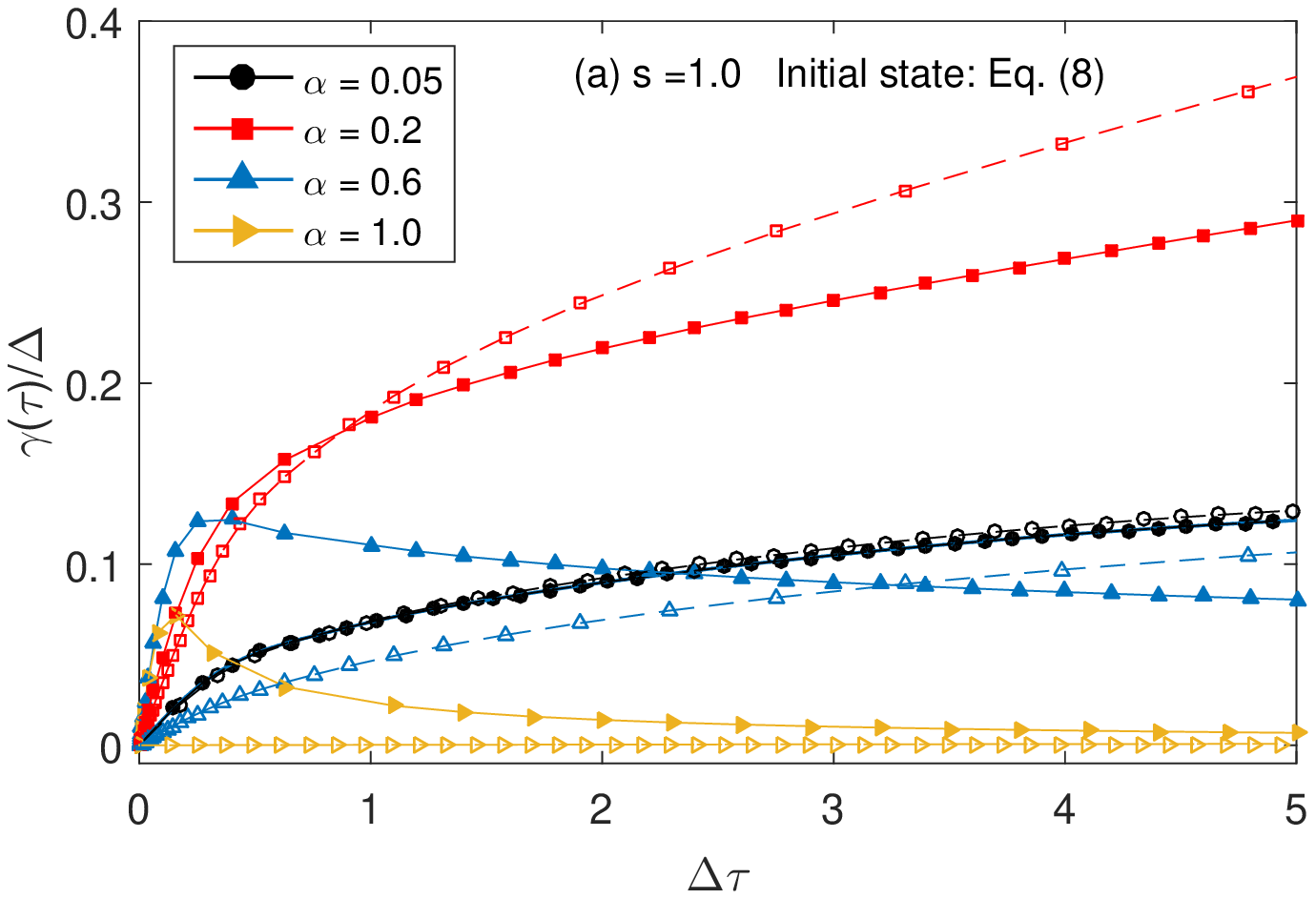} %
\includegraphics[scale=0.6]{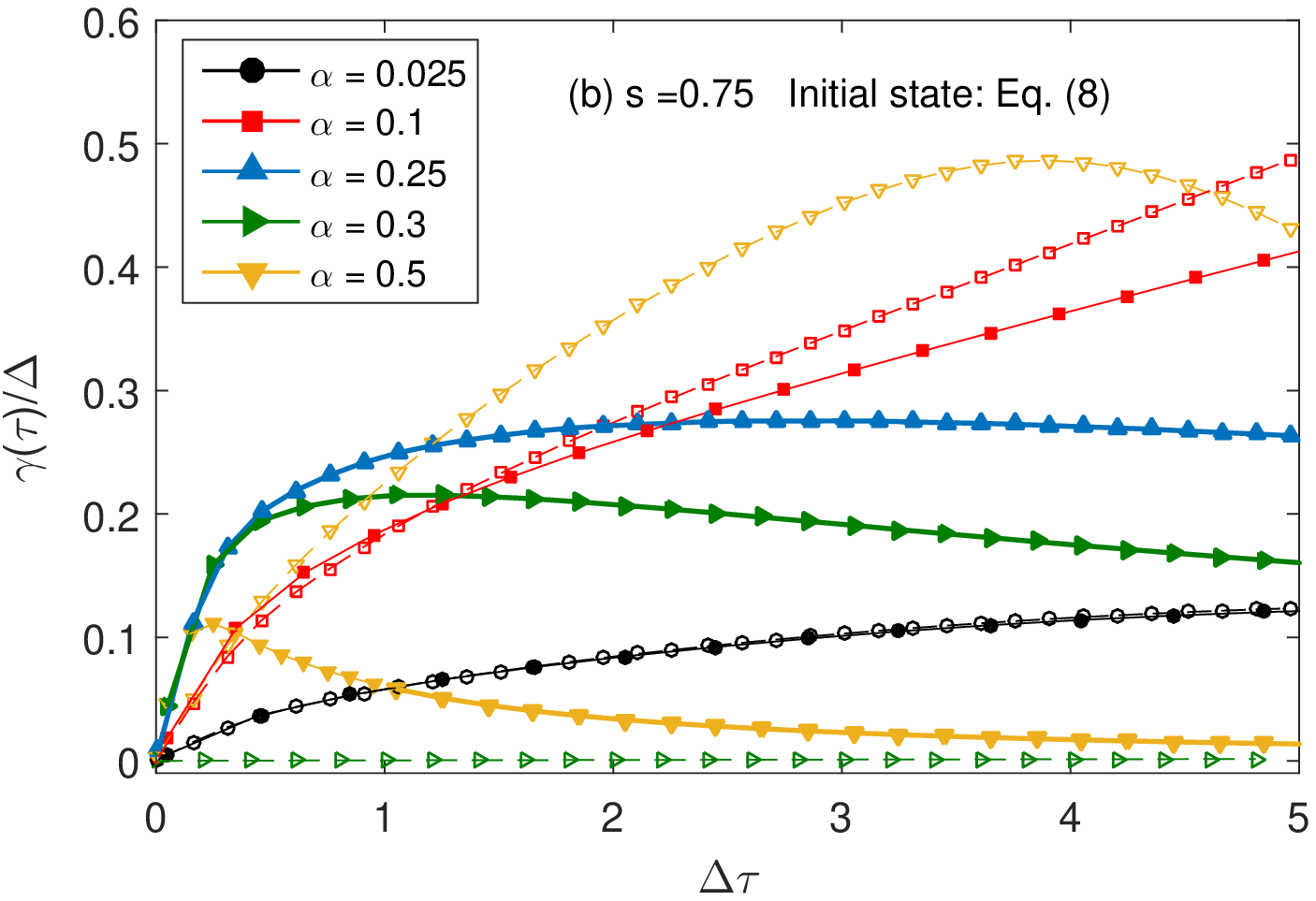}
\caption{(Color online) Zeno decay rate from  physical bath initial states obtained by TDVP (solid lines with filled symbols) and UT (dashed lines with open symbols)
methods for (a)  $s=1$ and (b) $s=0.75$. Specifically, thick solid lines denote the TDVP
results  for coupling strengths above the critical point  in the ground state.}
\label{Fig1}
\end{figure}

In this section, we measure the whole system by projecting the total wave
function onto the initial state. It means that we do not consider the
system-environment correlation effects in the measurement. Two initial
states are considered:

\textsl{Physical bath initial state}: The Zeno effect is easily studied
within the UT approach~\cite{Zheng} for the physical bath initial state (\ref%
{initial_2}). So we perform the TDVP study with the same initial state. The
comparison between the TDVP and UT method is presented in Fig. \ref{Fig1} for
bath exponent $s=1$ and $0.75$. As expected, the two methods agree well in
the weak coupling regime [e.g. $\alpha =0.05,s=1$ and $\alpha =0.025,s=0.75$%
\ ], since the counter-rotating terms play only a little role in this
regime, and the partial consideration of their effects in the UT study is a
very good approximation. \ It should be noted that the decay rate increases
monotonically with the increase of the time interval of measurements in the
weak coupling regime. This monotonicity can be regarded as a new definition
of the Zeno effect ~\cite{Lizuain}, since the evolution of the system is
slowed down by frequent measurement. The QAZE can be defined in an analogous
manner as the nonmonotonicity of $\gamma (\tau )$, i.e. the decay rate
increases first and then decreases again with the measurement time interval.
This definition retains the core physical content of QZE and QAZE, namely
that frequent measurement either slows down or accelerates the evolution.
Moreover, it avoids calculating the decay rate for the infinitely long
measurement interval which can be hardly obtained accurately by numerical
methods if no analytical solutions are available, such as in the present
model. 
Throughout this paper, we define QZE and QAZE by the functional behavior of $%
\gamma (\tau )$.

When the coupling strength increases, the UT result obviously deviates from
the numerically exact TDVP method. Interestingly, a non-monotonic behavior
is indeed found for strong coupling in the TDVP curves, which is never
exhibited by the UT curves indicated by dashed lines in Fig. \ref{Fig1}. The
TDVP method predicts a transition from QZE to QAZE for $\alpha \geq 0.6$
around $\Delta \tau =0.2$ and Ohmic bath. Note that UT fails to capture this
feature and only shows the QZE in all coupling regime ~\cite{Zheng}.
 It was demonstrated in Ref. ~\cite{Zach} that  the dynamics in the strong coupling regime
is considerably affected by the multi-photon process described by higher
order non-RWA terms  while the UT method neglects most of these
higher-order terms. Note also that the  significant differences
between RWA and non-RWA in the strong coupling regime was reported in
Ref. ~\cite{Prior2}. It is just these higher-order  non-RWA terms that are
responsible for the QZE-QAZE transition at the strong coupling.

\textsl{Bare bath initial state}: Starting the evolution from the widely
used state (\ref{initial_1}), the QZE-QAZE transition is always observed
both for the Ohmic and sub-Ohmic baths, as shown by solid lines in Fig. \ref%
{measure_qubit}, for all coupling strengths.  The previous RWA results, only applicable at extremely weak coupling,
also support these findings.

We conclude that the presence of the QZE-QAZE transition is highly dependent
on the choice of the initial state. The UT study for the physical bath
initial state can only be applied for weak coupling.  In the present exact
study, we find that this suppression of QAZE is an artefact of the UT
approximation and the QAZE reappears when the coupling strength increases.
On the other hand, for the often-used bare bath initial state, the
QZE-QAZE transition always obtains at arbitrary coupling.

\begin{figure}[hpt]
\includegraphics[scale=0.6]{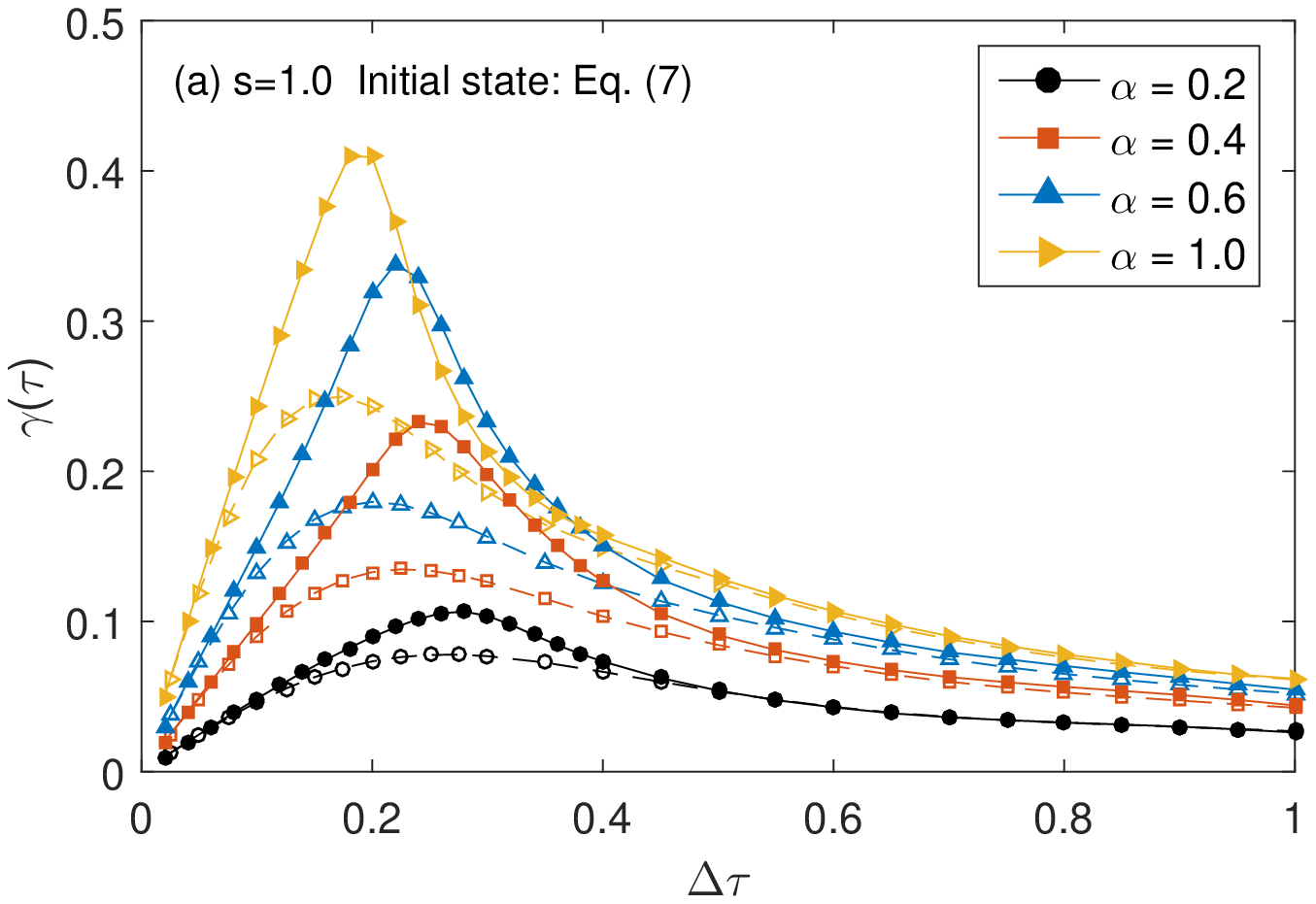} %
\includegraphics[scale=0.6]{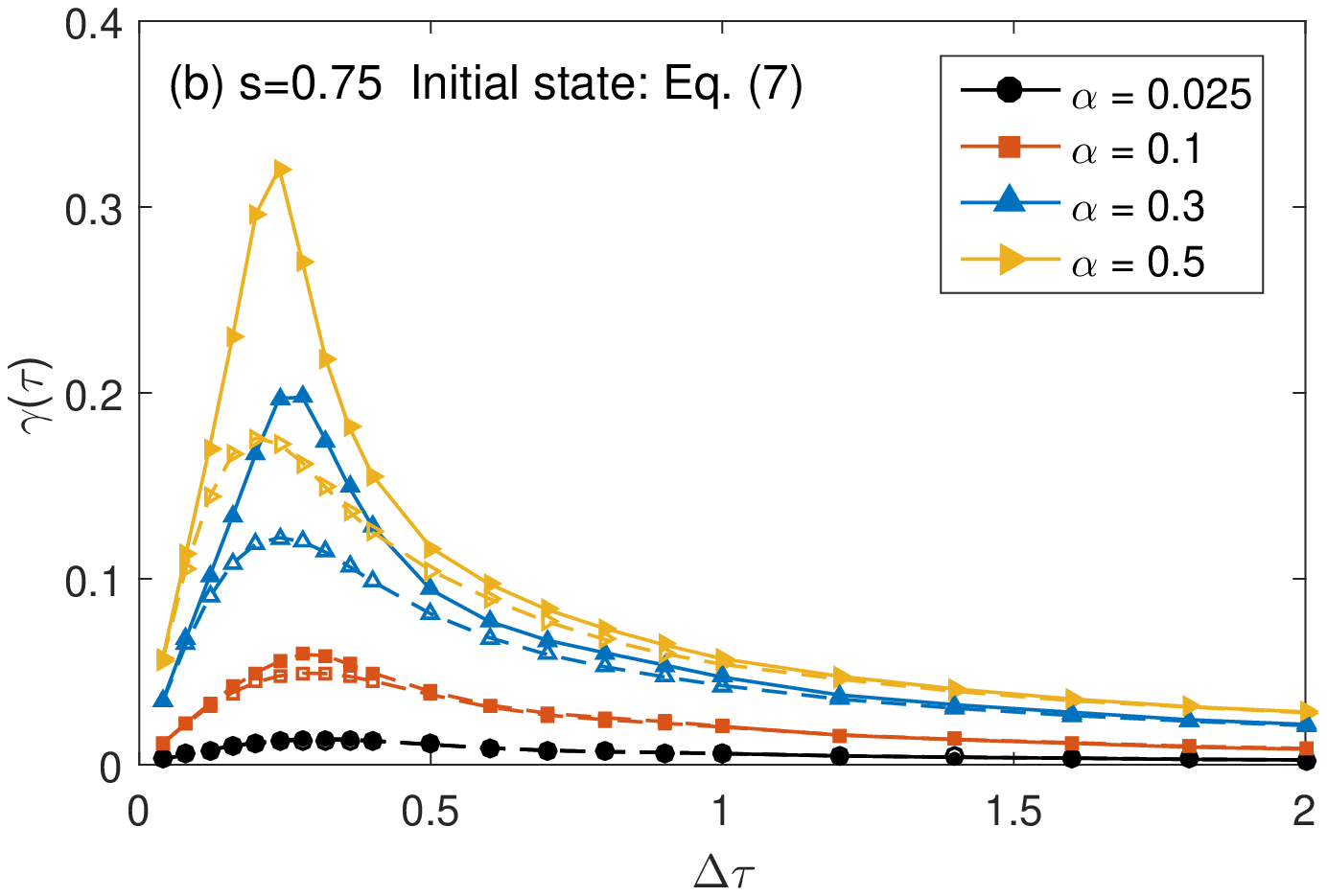}
\caption{(Color online) The decay rate from the bare bath initial states obtained by TDVP  for (a) $s=1$ and (b) $s=0.75$. Results with  the previous measurement scheme are indicated by  lines with open  symbols and those with the new measurement scheme by lines with filled symbols. }
\label{measure_qubit}
\end{figure}

In the spin-boson model, it is well known that the second-order quantum
phase transition from delocalization to localization occurs for sub-Ohmic
bath~\cite{QPT}.  While for the
Ohmic bath, the delocalized-localized quantum phase transition is of
Kosterlitz-Thouless type ~\cite{Leggett}. For the parameter used  in this paper, the critical coupling strengths are around $\alpha_c=1$ for $s=1$  and $0.295$ for $s=0.75$~\cite{QPT}.  From Fig. \ref{Fig1} we observe that the onset of the QZE-QAZE transition takes place below the corresponding critical points for both Ohmic and sub-Ohmic baths,  revealing that the presence of the localized, or
delocalized phase does not affect the QZE-QAZE transition.
In the study of Zeno effect, the incoherent decay of excited states
towards the ground state is highly nontrivial. The  reason that these ground state delocalization to localization transitions do not  affect the
the QZE-QAZE transition is worthy of further study. The preliminary understanding may be the following. The QZE and QAZE are very sensitive to the earlier dynamics, the  state evolving at the earlier stage is far away from the final ground-state. So the ground-state phase has no dominant  effect on the earlier state.  Even after  the long time evolution, the ground state have no essential effect. Above the critical point, the two states of the qubit are present with equal probability, so there is no influence at all. Below the critical point, a doubly degenerate localized phase is formed, the preferred spin state only depends on the initial
state, which is however independent of the coupling strength.

\subsection{Measuring the qubit only}

In the previous section, the measurement resets both the system and
environment to the total initial state. As explained in the Introduction,
this assumption is reasonable only if the coupling strength is weak and the
measurement is performed with high frequency. In the general case, each
measurement projects only the system (here the qubit) onto its initial
state, while the environment still evolves and never falls back to its
initial state upon measurement.
Here we describe a more general scheme suited to this case for the widely
used bare bath initial state (\ref{initial_1}).

After unitary evolution in a period of time $t$, the state can be written as
\begin{equation*}
|\psi (t)\rangle =C_{1}|\uparrow \rangle \otimes |\phi _{\text{env}%
}^{1}\rangle +C_{2}|\downarrow \rangle \otimes |\phi _{\text{env}}^{2}\rangle
\end{equation*}%
where $|\phi _{\text{env}}^{i,=1,2}\rangle $ is the environmental state of
photons. Measuring the qubit only (and projecting it by assumption onto its
initial spin-up state) will lead to the new initial state:
\begin{equation*}
|\psi (t)\rangle _{new}=|\uparrow \rangle \otimes |\phi _{\text{env}%
}^{1}\rangle .
\end{equation*}%
Note that the environmental part is always different from the initial photon
vacuum after the measurement due to the back-reaction of the system to the
environment, no matter how close they may be.

The decay rates for both the previous and the present  measurement scheme
are shown in Fig. \ref{measure_qubit} for exponents $s=1$ and $s=0.75$. It
is interesting that the decay rates differ markedly only for intermediate
time intervals. The qualitative nature of the QZE-QAZE transition remains
unchanged for all coupling strengths.

\begin{figure}[tph]
\includegraphics[scale=0.7]{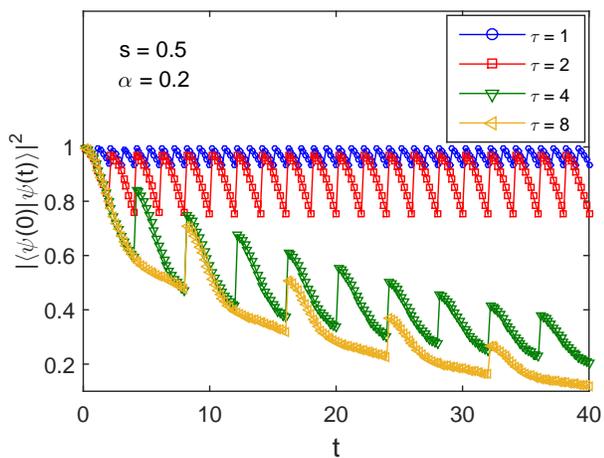}
\caption{(Color online) Evolution of the fidelity between current state and initial state
for different measurement time intervals in a sub-Ohmic bath. The initial
state is chosen as the bare bath initial state(\ref{initial_1}). }
\label{fidelity}
\end{figure}

To explain the above findings qualitatively, we calculate the time
dependence of the fidelity between the current state and the initial state
in the new measurement scheme for the sub-Ohmic ($s=0.5$) spin-boson model
at strong coupling ($\alpha =0.2$). It is clearly seen from Fig. \ref%
{fidelity} that the fidelity depends on the time interval $\tau $.

For fast frequent measurements like $\tau=1\sim 2$, the time interval is so short
that the environment cannot evolve far from its initial\ state, although the
correlation between the system and environment is strong, and the fidelity
almost remains the same (unity) after each measurement.  As shown in Fig. \ref%
{Photnum} (a),  the bath has only less than one
photonic excitation for the whole range of frequencies during the measurement
interval. The evolution of the environment exhibits an obvious periodicity,
indicating that the system plus its environment return to the initial state
after each measurement. In this case, the measurement scheme of the previous
section is not qualitatively different from the present new one. It
follows that the previous assumption still works well for fast measurements.

\begin{figure}[tph]
\includegraphics[scale=0.5]{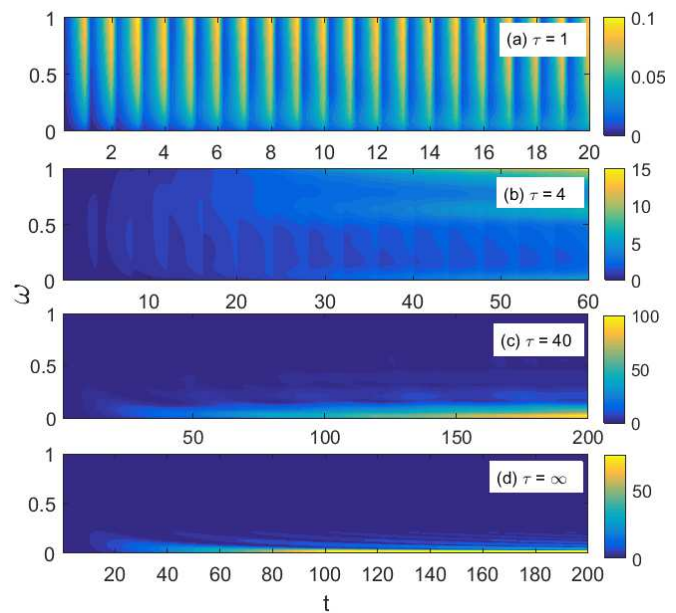}
\caption{ (Color online) Evolution of  numbers of photons  with different mode frequencies in the environment  for 4 typical measurement time
intervals when the  qubit is  measured only. The initial state is
chosen as the bare bath initial state~(\ref{initial_1}). $s=0.5$, $\alpha=0.2$.}
\label{Photnum}
\end{figure}

However, for slow measurements, $\tau=4\sim 8$, the fidelity dramatically
deviates from unity already after the second measurement, manifesting the
breakdown of the previous assumption. As demonstrated in Fig. \ref{Photnum}
(b),  the net excitations of the bath photonic numbers are accumulated through
successive measurement, especially for the high frequency part of the
environment. In this case, each measurement generates a new environmental
state which obvious evolves and deviates from its initial form due to the
strong correlation between the system and the environment. This new initial
state carries information from the previous evolution-measurement cycle into
the next one, inducing a non-Markovian effect to the whole repetitive
measurement process. The survival probability of each measurement turns out
to be time-dependent and finally enhances the decay rate.

When the measurement interval is large enough, exceeding the relaxation time (or
memory time of the bath), the qubit approximately relaxes to a quasi-steady
state which is independent from the initial state of the bath. It is exhibited in Fig. \ref%
{Photnum} (c) for  a relatively long time ($\tau =40$) that the
evolution of the bath during the measurement interval shows a gradually
excitations of the low frequency photons, similar to the one without
measurements shown in Fig. \ref{Photnum} (d). In this extremely slow measurements, the decay rate
remains unchanged with time, just demonstrated in the curves for large $\tau$ in Fig.~%
\ref{measure_qubit}.

\section{Conclusion}

In summary, we  study the Zeno effect in the open quantum system of a
qubit interacting with a bosonic bath by using a numerically exact method
based on the MPS. We find that the QZE-QAZE transition is highly dependent
on the initial state if measuring both spin (qubit) and environment, which
has been usually adopted in the literature. For the bare bath initial state,
the QZE-QAZE transition happens always, independent of the coupling
strength. For the physical bath initial state, the QZE-QAZE transition is
absent at weak coupling but reappears at strong coupling. Thus the previous
result of the UT-based study showing a suppression of QAZE ~\cite%
{Zheng08,suncp} has to be modified.

Furthermore, we consider a more realistic measurement process which only
projects the qubit to its initial state while the environment continues to
change. In the new measurement scheme, we find that QZE-QAZE transition
happens always, independent of the coupling strength, for the bare initial
state. In addition, measuring the qubit only leads to a faster decay rate
for intermediate measurement time intervals.

All the above statements holds for both Ohmic and sub-Ohmic baths. We did
not find any dependence of the above observations on the
coherent-incoherent crossover and/or the delocalized-localized phase
transition, indicating that these effects qualitatively play no role for the
QZE to QAZE transition.

Note added: Recently, we became aware of a  paper~\cite{linhq}
where the quantum Zeno and anti-Zeno effects were studied in a spin-boson
model with Lorentzian-like spectrum where the exact hierarchical
equations-of-motion approach can be applied. For the physical bath initial
state, multiple Zeno-to-anti-Zeno crossovers were observed, possibly due to
the oscillation of the population dynamics in the Lorentzian bath as shown
in Fig.~1 of Ref. ~\cite{linhq}. The Lorentzian bath is usually used to describe a
bad cavity, so the population dynamics is similar to the case of single
(lossy) cavity, like the quantum Rabi model. In contrast, for Ohmic and
sub-Ohmic baths we find here that the QZE-QAZE transition usually
occurs only once.

\textbf{ACKNOWLEDGEMENTS} This work was supported by th National Natural
Science Foundation of China under Grants No. 11674285 and No. 11474256. We thank D. Braak for  critically reading this manuscript  and many helpful suggestions.  We also acknowledge useful discussions with C. Wang.

$^{*}$ Email:qhchen@zju.edu.cn



\end{document}